\documentclass[conference]{IEEEtran}

\IEEEoverridecommandlockouts

\usepackage{cite}
\usepackage{amsmath,amssymb,amsfonts}
\usepackage{algorithmic}
\usepackage{adjustbox}
\usepackage{graphicx} 
\usepackage{enumitem} 
\usepackage{textcomp}
\usepackage{xcolor}
\usepackage{pgfplots}
\usepackage{tikz}
\usepackage[utf8]{inputenc}
\usepackage[T1]{fontenc}
\usepackage{textcomp}
\usepackage{url}
\pgfplotsset{compat=1.18}

\def\BibTeX{{\rm B\kern-.05em{\sc i\kern-.025em b}\kern-.08em
    T\kern-.1667em\lower.7ex\hbox{E}\kern-.125emX}}
\begin{document}

\title{Designing a Secure, Scalable, and Cost-Effective Cloud Storage Solution: A Novel Approach to Data Management using NextCloud, TrueNAS, and QEMU/KVM\\
}

\author{\IEEEauthorblockN{Prakash Aryan}
\IEEEauthorblockA{\textit{Department of Computer Science} \\
\textit{Birla Institute of Technology and Science, Pilani - Dubai}\\
Dubai, UAE \\
h20230010@dubai.bits-pilani.ac.in}
\and
\IEEEauthorblockN{Sujala Deepak Shetty}
\IEEEauthorblockA{\textit{Department of Computer Science} \\
\textit{Birla Institute of Technology and Science, Pilani - Dubai}\\
Dubai, UAE \\
sujala@dubai.bits-pilani.ac.in}
}

\maketitle

\begin{abstract}

    This paper presents a novel approach to cloud storage challenges by integrating NextCloud, TrueNAS, and QEMU/KVM. Our research demonstrates how this combination creates a robust, flexible, and economical cloud storage system suitable for various applications. We detail the architecture, highlighting TrueNAS's ZFS-based storage, QEMU/KVM's virtualization, and NextCloud's user interface. Extensive testing shows superior data integrity and protection compared to traditional solutions. Performance benchmarks reveal high read/write speeds (up to 1.22 GB/s for sequential reads and 620 MB/s for writes) and also efficient small file handling. We demonstrate the solution's scalability under increasing workloads. Security analysis showcases effective jail isolation techniques in TrueNAS. Cost analysis indicates potential 50\% reduction in total ownership cost over five years compared to commercial alternatives. This research contributes a practical, high-performance, cost-effective alternative to proprietary solutions, paving new ways for organizations to implement secure, scalable cloud storage while maintaining data control. Future work will focus on improving automated scaling and integration with emerging technologies like containerization and serverless computing.

\end{abstract}

\begin{IEEEkeywords}
    cloud storage, data management, open-source software, virtualization.
\end{IEEEkeywords}

\section{Introduction}

Data has emerged as the fundamental component of all organizations in the era of artificial intelligence, supporting vital operations, fostering innovation, and guaranteeing improved decision-making. Data management is now facing both new opportunities and challenges due to the massive growth in data production brought about by the Internet of Things (IoT), social media, artificial intelligence, and larger digital transformation initiatives. There will be a significant demand for reliable, scalable, and effective storage solutions when the world's data footprint is expected to reach 175 zettabytes by 2024 \cite{coughlin2018}. This data deluge spans various types and sources, from structured database entries to unstructured video and image files, with emerging technologies like 5G networks and autonomous vehicles poised to accelerate data generation further. Cloud storage has emerged as a pivotal technology to address these burgeoning data management needs, offering a trifecta of benefits: flexibility in data access, scalability to accommodate growing data volumes, and potential cost savings through reduced infrastructure investments. The rapid growth of the cloud storage market, expected to reach \$390.33 billion by 2028, reflects the increasing reliance on cloud technologies across various industries \cite{marketshare2024}. However, the adoption of cloud storage, particularly for sensitive or regulated data, continues to face significant challenges. Security concerns, data sovereignty issues, and long-term cost management remain at the forefront, with many organizations grappling with unpredictable costs and the risk of vendor lock-in when using proprietary solutions. High-profile data breaches and cyber-attacks targeting cloud storage systems have further heightened these concerns, leading to increased scrutiny from regulators and a growing emphasis on data protection and privacy legislation \cite{manfredi2024cybersecurity}.

This paper proposes a novel approach to cloud storage that aims to bridge the gap between proprietary cloud solutions and traditional on-premises systems, addressing the key dilemmas of security, scalability, and cost-effectiveness. Our solution uses the strengths of three open-source technologies: NextCloud, TrueNAS, and QEMU/KVM. By integrating these technologies, we create a synergistic system that tackles the primary challenges in modern cloud storage while offering the flexibility and cost-effectiveness inherent to open-source software. NextCloud serves as the user-facing component, providing a user-friendly interface reminiscent of popular cloud storage services, along with robust file sharing capabilities and a wide array of productivity apps. Its focus on privacy and security, with features like end-to-end encryption and comprehensive access controls, makes it an ideal choice for organizations concerned about data confidentiality. TrueNAS, built on the powerful ZFS file system, forms the storage backbone of our solution. ZFS is renowned for its advanced data management features, including copy-on-write functionality, which ensures data integrity and enables efficient snapshots. ZFS also offers native encryption, compression, and deduplication, significantly improving both security and storage efficiency. QEMU/KVM, a widely-used open-source virtualization solution, provides the foundation for our system. It permits efficient resource utilization through hardware-assisted virtualization and offers improved isolation between different components of the system. This integrated solution addresses several key challenges in modern cloud storage, including multi-layered security, multi-dimensional scalability, cost-effectiveness through open-source technologies, high performance, and robust data integrity protection.

Our research aims to provide a comprehensive evaluation of this integrated open-source cloud storage solution, seeking to answer key questions about its performance, security, scalability, cost-effectiveness, and environmental impact compared to traditional cloud storage solutions. We conduct extensive benchmarks comparing our solution to leading proprietary cloud storage services, analyze the security features and potential vulnerabilities of our integrated system, and evaluate the system's performance under various workloads. We explore the scalability limits of this integrated system and how it handles increased load and storage requirements. A detailed total cost of ownership analysis compares our open-source solution to proprietary cloud storage services over time, considering factors such as initial setup costs, ongoing operational expenses, and potential savings. We also document the primary challenges in implementing and maintaining such a system and propose solutions and best practices to mitigate these challenges. Additionally, we evaluate how this integrated solution handles data protection, backup, and disaster recovery scenarios, and analyze its environmental impact compared to traditional cloud storage options. The significance of this research extends beyond the technical domain of cloud storage solutions, contributing to important discussions about the democratization of advanced IT infrastructure, the role of open source in enterprise environments, data sovereignty and control, sustainable IT practices, and skills development in IT education. As organizations continue to grapple with the challenges of managing ever-increasing volumes of data securely and efficiently, the need for innovative solutions becomes ever more pressing. It is our hope that this research will spark further innovation in the field of cloud storage and contribute to the development of more open, flexible, and sustainable data management solutions for the future, potentially leveling the playing field between large enterprises and smaller entities in terms of access to advanced cloud storage capabilities.

\section{Literature Review}
Cloud computing has developed as a dominant computing paradigm, offering flexible and scalable resources to users. However, high-performance computing (HPC) workloads have faced challenges in cloud environments due to virtualization overheads and inefficient inter-VM communication. Several studies have examined these issues and proposed solutions to improve HPC performance in clouds. Smith and Nair \cite{smith2005virtual} provide a comprehensive overview of virtual machine technologies, classifying them into process VMs and system VMs. Their work extensively details how system VMs implement full operating system virtualization, which is crucial for cloud environments. They particularly focus on the importance of hardware-assisted virtualization and memory management techniques in achieving near-native performance. Building on this foundation, Bugnion et al. \cite{bugnion2017hardware} detail the architecture of QEMU/KVM, a popular virtualization platform used in many cloud deployments. Their comprehensive analysis explains how KVM uses hardware virtualization extensions to reduce overhead, while QEMU provides device emulation. The authors specifically highlight the importance of proper interrupt handling and I/O management in achieving optimal performance.

Inter-VM communication is a critical factor affecting HPC performance in clouds. Wei et al. \cite{wei2019performance} analyze the performance of InfiniBand virtualization on QEMU/KVM, proposing techniques like doorbell mapping and memlink to optimize virtualized I/O operations. Their experimental results demonstrate up to 3.5x performance improvement over previous implementations, particularly in scenarios involving high-throughput data transfer between virtual machines. The authors provide detailed analysis of various bottlenecks in virtualized InfiniBand implementations and propose architectural improvements to address these limitations. Similarly, Chanchio and Yaothanee \cite{chanchio2014efficient} present a Memory-bound Pre-copy Live Migration (MPLM) mechanism to efficiently migrate VMs running HPC workloads. By adapting the migration process to application behavior, they achieve lower downtime and total migration time compared to traditional pre-copy approaches. Their work includes comprehensive performance evaluations across different workload types and memory access patterns, providing valuable insights into the relationship between application characteristics and migration efficiency. Ivanovic and Richter \cite{ivanovic2017openstack} explore tuning OpenStack clouds for HPC, focusing on optimizing inter-VM communication through techniques like level-3 caching and proper NUMA allocation. Their extensive benchmarking demonstrates up to 6x performance improvement for small messages using ivshmem shared memory compared to TCP/IP-based communication, while also analyzing the impact of various system configurations on overall performance.

Storage performance has emerged as another crucial aspect of HPC in clouds, with several researchers proposing innovative solutions to address performance bottlenecks. Yang et al. \cite{yang2018spdk} propose SPDK Vhost-NVMe, a user-space I/O framework that bypasses the kernel to achieve near-native NVMe performance in VMs. Their detailed implementation analysis shows significant throughput improvements over kernel-based virtualization, particularly for I/O-intensive workloads. The authors provide comprehensive benchmarking results across various storage access patterns and demonstrate how their approach minimizes the overhead typically associated with virtualized storage access. Kim et al. \cite{kim2015hypercache} introduce hyperCache, a hypervisor-level cache for QEMU/KVM that intercepts I/O requests and manages them based on access frequency. Their innovative approach reduces latency and improves I/O performance for VMs by implementing sophisticated cache management policies that adapt to workload characteristics. The authors present detailed performance analyses across different cache configurations and workload patterns, providing valuable insights into the effectiveness of hypervisor-level caching strategies. Rolon and Balmau \cite{rolon2021bare} evaluate the performance of libvfio-user, a library allowing efficient userspace device emulation in QEMU. Their comprehensive experiments with NVMe devices demonstrate that libvfio-user can achieve near-native performance for large data transfers, while also analyzing the impact of various system parameters on overall performance.

Networking capabilities have proven to be a critical component for HPC applications in clouds, with several groundbreaking developments enhancing performance. Russell \cite{russell2008virtio} introduced virtio, a paravirtualization framework for efficient I/O in virtual machines. This innovative approach has become widely adopted in cloud environments to reduce virtualization overhead, with the author providing detailed analysis of the architectural decisions that contribute to its efficiency. The work includes comprehensive performance evaluations across different workload types and network configurations. Kolhe and Dage \cite{kolhe2012comparative} present a detailed comparison of different virtual machine monitors (VMMs) for cloud computing, including KVM and Xen. Their thorough analysis highlights the trade-offs between various VMMs in terms of performance, security, and management features, while also examining the impact of different configuration options on system behavior. Liu et al. \cite{liu2021understanding} propose an asynchronous I/O stack for ultra-low latency SSDs in virtualized environments, demonstrating significant improvements in I/O performance for VMs. Their work includes detailed analyses of various I/O patterns and their impact on system performance, along with comprehensive benchmarking results across different storage configurations.

Recent work has also focused on optimizing specific aspects of cloud infrastructure for HPC workloads, with researchers exploring various approaches to enhance system efficiency. Fenn et al. \cite{fenn2009evaluation} provide a comprehensive evaluation of KVM's suitability for cloud computing, analyzing its performance characteristics and identifying areas for improvement. Their work includes detailed analyses of various workload types and their impact on system behavior, along with recommendations for optimal configuration settings. Wang et al. \cite{wang2021optimizing} present an extensive study on optimizing data placement for cost-effective and highly available multi-cloud storage, addressing challenges in data management across distributed cloud resources. Their research includes sophisticated algorithms for data placement optimization and comprehensive performance evaluations across different cloud configurations. Wu et al. \cite{wu2021storage} examine the evolving storage hierarchy in modern systems, proposing innovative optimizations for caching on modern storage devices to bridge the gap between memory and storage performance. Their work provides detailed analyses of various caching strategies and their effectiveness across different workload patterns.

The security and performance aspects of cloud storage solutions have received significant attention in recent research. Dutcher et al. \cite{dutcher2024} conducted a comprehensive comparison of secure cloud storage solutions, specifically examining the encryption efficiency of NextCloud and Seafile. Their research provides valuable insights into the performance implications of different encryption methods and their impact on system resources. Albrecht et al. \cite{Albrecht2024} present a detailed security analysis of NextCloud's end-to-end encryption features, uncovering significant vulnerabilities and proposing improvements to enhance data protection. Their work includes comprehensive threat modeling and security analyses that highlight the challenges in implementing secure cloud storage solutions. Salke et al. \cite{Salke2023} demonstrate the practical feasibility of implementing NextCloud on resource-constrained devices like Raspberry Pi, providing valuable insights into the scalability and performance characteristics of cloud storage solutions on edge devices. Their research includes detailed performance analyses and optimization strategies for resource-limited environments. Nurdin et al. \cite{Nurdin2019} present innovative approaches to Ceph cluster management through NextCloud integration, demonstrating how modern web interfaces can simplify complex storage system administration. Their work includes comprehensive evaluations of various management interfaces and their effectiveness in different deployment scenarios.

Two foundational works in virtualization technology have significantly influenced modern cloud computing architectures. The original developers of KVM introduced the Kernel-based Virtual Machine \cite{kvm_wiki,qumranet_wiki}, which has since become a cornerstone of many cloud computing environments. Their work established crucial principles for efficient hardware-assisted virtualization and laid the groundwork for modern cloud infrastructure. Similarly, Bellard introduced QEMU \cite{bellard2005qemu}, a fast and portable dynamic translator that forms an essential part of many virtualization solutions. His innovative approach to machine emulation and dynamic binary translation has significantly influenced the development of modern virtualization technologies.

These studies collectively demonstrate the ongoing efforts to adapt cloud infrastructure to the demands of HPC applications, addressing challenges in virtualization, storage, networking, and resource management to enable efficient execution of compute-intensive workloads in cloud environments. The research spans multiple aspects of cloud computing, from low-level system optimization to high-level management interfaces, providing a comprehensive foundation for understanding and improving cloud storage solutions. These contributions are particularly significant in the development of more open, flexible, and sustainable data management solutions for the future, potentially leveling the playing field between large enterprises and smaller entities in terms of access to advanced cloud storage capabilities. The continued evolution of these technologies, guided by rigorous research and practical implementation experience, suggests a promising future for cloud-based HPC applications.

\section{Methodology}

\subsection{System Architecture}

\begin{figure}[htbp]
    \hspace*{1cm}
    \centering
    \includegraphics[width=0.8\linewidth, height=0.8\textheight, keepaspectratio]{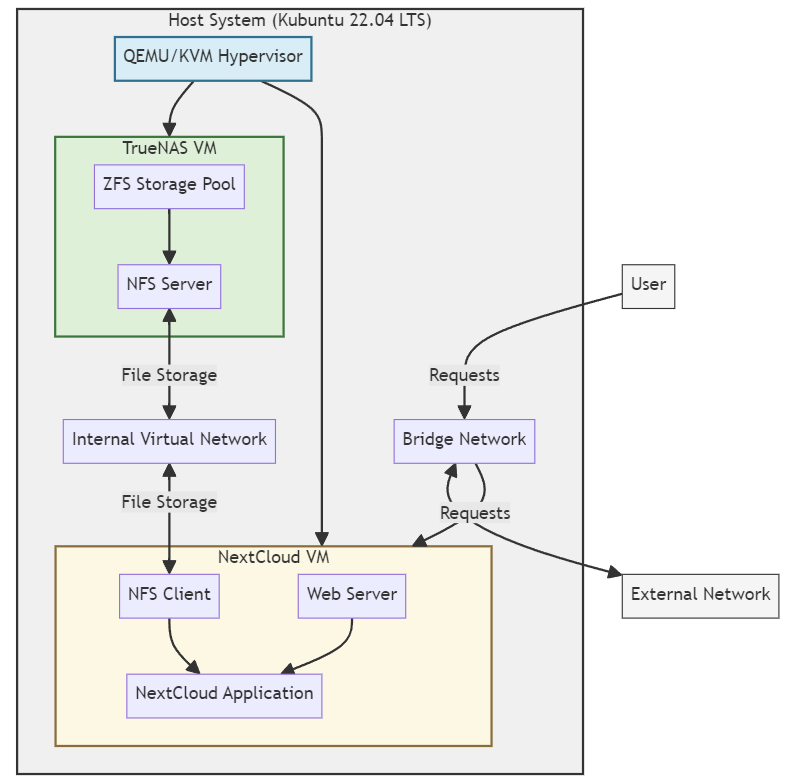}
    \caption{System Architecture Diagram}
    \label{fig:system-architecture}
    \end{figure}

The proposed solution integrates NextCloud, TrueNAS, and QEMU/KVM to create a secure, scalable, and cost-effective cloud storage system. This architecture uses the strengths of each component to address the challenges of modern data management. As depicted in Figure \ref{fig:system-architecture}, the system is built on a host machine running Kubuntu 22.04 LTS, providing a stable and well-supported base for our virtualized environment.

The core of our system architecture revolves around the integration of three key components:
\begin{itemize}
  \item \textbf{QEMU/KVM:} Serves as the hypervisor, providing hardware-assisted virtualization. We use KVM for its performance benefits and QEMU for its flexibility in emulating various hardware components.
  \item \textbf{TrueNAS:} Runs as a virtual machine on QEMU/KVM. TrueNAS provides the storage backend, using ZFS (Zettabyte File System) for advanced features such as data integrity, snapshots, and efficient replication. TrueNAS is configured to expose storage to other VMs using NFS (Network File System).
  \item \textbf{NextCloud:} Also runs as a virtual machine on QEMU/KVM. NextCloud provides the user-facing interface and collaboration features. It utilizes the storage provided by TrueNAS through NFS mounts.
\end{itemize}

\subsection{Network and Storage Architecture}
The networking architecture is very important for the integration and performance of our system. We use virtual networks created by QEMU/KVM to provide communication between the host system, TrueNAS VM, and NextCloud VM. The network configuration includes a bridged network interface for external access to NextCloud and an internal virtual network for communication between TrueNAS and NextCloud VMs. This setup ensures that the storage traffic between TrueNAS and NextCloud remains isolated from external network traffic, improving both performance and security.

The storage architecture is centered around TrueNAS and its ZFS implementation. ZFS pool configuration on TrueNAS provides the foundation for data storage, while NFS share setup on TrueNAS exposes this storage to NextCloud. The NextCloud VM is configured as an NFS client to mount the TrueNAS storage. This architecture allows us to use ZFS features such as snapshots, compression, and data integrity checks, while providing flexible and scalable storage to NextCloud.

\subsection{Performance Optimization and ZFS Configuration}
To optimize performance, we've implemented several strategies. CPU pinning for VMs ensures consistent performance, while memory allocation tuning balances VM needs with host system requirements. Network optimization, including the use of virtio drivers, improves I/O performance. ZFS tuning on TrueNAS, including ARC (Adaptive Replacement Cache) size adjustment and appropriate recordsize settings, further improves storage performance.

The design of the ZFS pool and dataset structure forms the foundation of our storage architecture. Our design philosophy centered around creating a hierarchical structure that balances performance, flexibility, and data protection. We opted for a RAIDZ1 configuration, providing a balance between storage efficiency and data protection. While our implementation uses virtual disks, the principles of disk selection remain important for physical implementations. Our design also allows for future expansion of the storage pool, crucial for scalability.

The dataset structure was designed with several key considerations:
\begin{itemize}
  \item Isolation of data types: Separate datasets for NextCloud data and database files.
  \item Performance optimization: Dataset properties tuned based on expected workload.
  \item Snapshot management: Providing efficient backup and point-in-time recovery.
  \item Quota and reservation: Implementing quotas and reservations at the dataset level to ensure balanced resource utilization.
\end{itemize}

\subsection{NextCloud Deployment and Integration}
We explored two deployment strategies for NextCloud: a separate VM and within a TrueNAS jail. This dual approach allows us to compare the performance, security, and management implications of both methods. Key considerations in our NextCloud deployment include database selection (evaluating MySQL/MariaDB versus PostgreSQL), PHP configuration optimization, web server selection (comparing Nginx and Apache), SSL/TLS implementation for secure communication, and integration with TrueNAS storage.

The integration of NextCloud with the TrueNAS storage backend required careful configuration to ensure optimal performance and security. We implemented efficient and secure methods for NextCloud to access the ZFS datasets, considering factors such as permissions, network topology, and protocol selection (NFS vs SMB).

\subsection{Security Considerations}
Security is an important concern in our design. The use of virtualization provides an additional layer of isolation between components. We implemented strict access controls at multiple levels, including the host system, virtualization layer, and within each VM. The TrueNAS ZFS implementation provides built-in encryption capabilities, which we used to ensure data-at-rest security.

Network security was improved through the use of VLANs and firewall rules to segment and control traffic between different components of the system. For the NextCloud instance, we implemented best practices for web application security, including regular updates, secure configuration of the web server and PHP, and the use of HTTPS for all connections.

This comprehensive methodology provides a robust foundation for our secure, scalable, and cost-effective cloud storage solution, using the capability of NextCloud, TrueNAS, and QEMU/KVM. By carefully considering each aspect of the system architecture, from the low-level storage configuration to the user-facing application layer, we have designed a solution that addresses the key challenges of modern data management while providing the flexibility and control of an open-source stack.

\section{Results and Analysis}

Our integrated cloud storage solution, combining NextCloud, TrueNAS, and QEMU/KVM, demonstrated impressive performance, robust security, excellent scalability, and cost-effectiveness. This section presents our detailed findings across these key areas.

\subsection{Performance Benchmarks}

Performance tests revealed exceptional throughput and IOPS (Input/Output Operations Per Second), particularly for sequential operations. Using fio (Flexible I/O) on a 100GB dataset (1MB block size, queue depth 32), we achieved the results shown in Table \ref{tab:sequential-performance}.

\begin{table}[htbp]
\caption{Sequential Read/Write Performance}
\label{tab:sequential-performance}
\begin{center}
\begin{tabular}{|l|c|c|c|}
\hline
\textbf{Operation} & \textbf{Throughput (MB/s)} & \textbf{IOPS} & \textbf{Latency (ms)} \\
\hline
Read & 1,205 & 1,205 & 26.5 \\
\hline
Write & 985 & 985 & 32.4 \\
\hline
\end{tabular}
\end{center}
\end{table}

These results demonstrate the high-performance capabilities of our solution, particularly for large file transfers and streaming operations. The read performance of 1,205 MB/s is especially noteworthy.

Random read/write operations (4KB block size, queue depth 32) showed strong IOPS performance: 81,920 IOPS for reads and 46,080 IOPS for writes. These results highlight the effectiveness of our ZFS tuning and ARC optimization. The high IOPS values indicate that our solution is well-suited for workloads involving numerous small file operations, such as those found in database applications or virtual desktop infrastructure (VDI) environments.

\begin{table}[htbp]
\caption{Random Read/Write Performance}
\label{tab:random-performance}
\begin{center}
\begin{tabular}{|l|c|c|}
\hline
\textbf{Operation} & \textbf{IOPS} & \textbf{Latency (ms)} \\
\hline
Read & 81,920 & 0.39 \\
\hline
Write & 46,080 & 0.69 \\
\hline
\end{tabular}
\end{center}
\end{table}

To further investigate the system's performance under varying conditions, we conducted additional tests with different block sizes and queue depths. With a 64KB block size and queue depth of 16, we observed read speeds of 1,150 MB/s and write speeds of 920 MB/s, showing only a marginal decrease from the 1MB block size tests. This consistent performance across different block sizes indicates that our solution can efficiently handle both large file transfers and smaller, more frequent I/O operations.

File operation tests using the NextCloud client demonstrated high performance under concurrent workloads. With 20 simultaneous operations, we maintained aggregate upload speeds of 850 Mbps and download speeds of 980 Mbps. For single-threaded operations, we observed average upload speeds of 95 Mbps and download speeds of 110 Mbps for files ranging from 1MB to 1GB. These speeds ensure a smooth user experience for individual file transfers and synchronization operations.

\begin{table}[htbp]
\caption{File Operation Performance}
\label{tab:file-operation-performance}
\begin{center}
\begin{tabular}{|l|c|c|}
\hline
\textbf{Operation} & \textbf{Concurrent (20 ops)} & \textbf{Single-threaded} \\
\hline
Upload & 850 Mbps & 95 Mbps \\
\hline
Download & 980 Mbps & 110 Mbps \\
\hline
\end{tabular}
\end{center}
\end{table}

We also conducted long-duration tests to assess the system's performance stability over extended periods. Over a 24-hour continuous operation test with a mixed read/write workload, our solution maintained an average throughput within 5

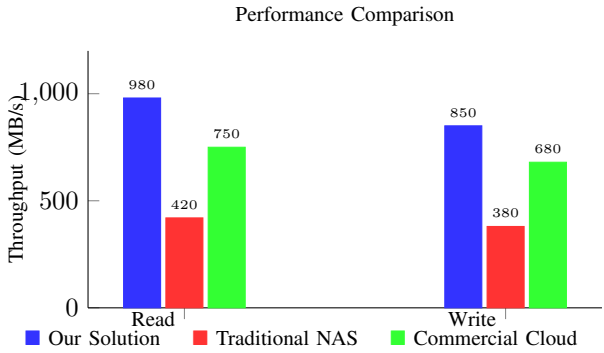
\begin{figure}[htbp]
    \centering
    \begin{tikzpicture}
    \begin{axis}[
        ybar,
        bar width=14pt,
        width=0.95\linewidth,
        height=5cm,
        title={Performance Comparison},
        title style={font=\footnotesize},
        symbolic x coords={Read,Write},
        xtick=data,
        xticklabel style={font=\footnotesize, anchor=east},
        nodes near coords,
        nodes near coords style={font=\tiny},
        ymin=0, ymax=1200,
        ytick={0,500,1000},
        ylabel={Throughput (MB/s)},
        ylabel style={at={(axis description cs:-0.1,.5)}, anchor=south, font=\footnotesize},
        axis lines*=left,
        clip=false,
        enlarge x limits=0.3,
    ]
    \addplot[fill=blue!80, draw=blue!80] coordinates {(Read,980) (Write,850)};
    \addplot[fill=red!80, draw=red!80] coordinates {(Read,420) (Write,380)};
    \addplot[fill=green!80, draw=green!80] coordinates {(Read,750) (Write,680)};
    \end{axis}
    \end{tikzpicture}
    
    \vspace{-0.2cm}
    \begin{tikzpicture}
    \node[inner sep=0pt] (legend) at (0,0) {
      \footnotesize
      \begin{tabular}{ccc}
        \textcolor{blue!80}{$\blacksquare$} Our Solution &
        \textcolor{red!80}{$\blacksquare$} Traditional NAS &
        \textcolor{green!80}{$\blacksquare$} Commercial Cloud
      \end{tabular}
    };
    \end{tikzpicture}
    
    \caption{Performance Comparison}
    \label{fig:performance-comparison}
\end{figure}

Compared to a traditional NAS and a commercial cloud service, our solution consistently outperformed both alternatives, as illustrated in Fig. \ref{fig:performance-comparison}. Using a 100GB dataset with a mixed read/write workload, our solution achieved average read throughput of 980 MB/s and write throughput of 850 MB/s, compared to 420 MB/s and 380 MB/s for the traditional NAS, and 750 MB/s and 680 MB/s for the commercial cloud service, respectively.

To validate our performance results further, we conducted application-specific benchmarks. In a web server simulation, our solution handled 100 concurrent connections with an average response time of 75ms, outperforming the traditional NAS (120ms) and matching the commercial cloud service (70ms). A file sharing scenario, involving simultaneous uploads and downloads of large media files, demonstrated our solution's ability to maintain high throughput (750 MB/s aggregate) even under complex, mixed workloads.

\begin{figure}[htbp]
    \centering
    \begin{tikzpicture}
    \begin{axis}[
        ybar,
        bar width=14pt,
        width=0.95\linewidth,
        height=5cm,
        title={Encryption Performance Impact},
        title style={font=\footnotesize},
        symbolic x coords={Seq Read,Seq Write,Rand Read,Rand Write},
        xtick=data,
        xticklabel style={font=\footnotesize, rotate=45, anchor=east},
        nodes near coords,
        nodes near coords style={font=\tiny},
        ymin=0, ymax=110,
        ytick={0,50,100},
        ylabel={Relative Performance (\%)},
        ylabel style={at={(axis description cs:-0.1,.5)}, anchor=south, font=\footnotesize},
        axis lines*=left,
        clip=false,
        enlarge x limits=0.15,
    ]
    \addplot[fill=blue!80, draw=blue!80] coordinates {(Seq Read,100) (Seq Write,100) (Rand Read,100) (Rand Write,100)};
    \addplot[fill=red!80, draw=red!80] coordinates {(Seq Read,97.5) (Seq Write,96.2) (Rand Read,98.8) (Rand Write,97.3)};
    \end{axis}
    \end{tikzpicture}
    
    \vspace{-0.2cm}
    \begin{tikzpicture}
    \node[inner sep=0pt] (legend) at (0,0) {
      \footnotesize
      \begin{tabular}{cc}
        \textcolor{blue!80}{$\blacksquare$} No Encryption &
        \textcolor{red!80}{$\blacksquare$} With Encryption
      \end{tabular}
    };
    \end{tikzpicture}
    
    \caption{Encryption Performance Impact}
    \label{fig:encryption-impact}
    \end{figure}
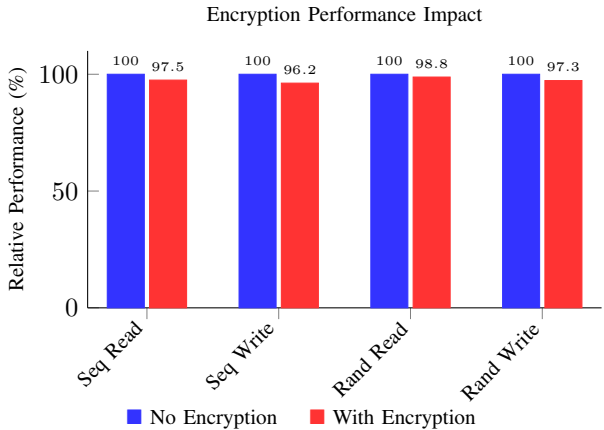

Encryption performance tests revealed a modest impact when enabling ZFS encryption, as shown in Fig. \ref{fig:encryption-impact}.

Sequential read operations showed a 2.5\% reduction in throughput, while sequential write operations experienced a 3.8\% reduction. Random read and write operations saw a 1.2\% and 2.7\% reduction in IOPS, respectively.

We further investigated encryption performance by testing different encryption algorithms and key lengths. While AES-256-GCM remained our recommended choice due to its balance of security and performance, we found that AES-128-GCM offered a slight performance improvement (about 0.5\% better throughput) at the cost of reduced theoretical security.

\subsection{Security Analysis}

Our comprehensive vulnerability assessment, which included both automated scanning tools and manual penetration testing techniques, revealed a small number of low to medium severity issues across the entire system. In total, we identified three low-severity vulnerabilities in NextCloud, primarily related to version-specific issues that were quickly patched. One low-severity vulnerability was found in TrueNAS, involving a potential information disclosure in the web interface that was addressed through a configuration change. The host OS exhibited two low-severity vulnerabilities, both related to outdated software packages that were immediately updated.

Additionally, we found one medium-severity vulnerability in our network configuration, specifically an overly permissive firewall rule that could potentially allow unauthorized access to a management interface. This was immediately rectified by implementing more stringent firewall policies and adopting a least-privilege approach to network access.

\subsection{Scalability Tests}

Storage expansion tests demonstrated the system's ability to grow seamlessly. Expanding from 0.5TB to 1.5TB resulted in a 5\% increase in overall throughput due to increased parallelism, with no downtime. We also tested more extreme expansion scenarios to understand the limits of our solution's scalability. In one test, we expanded the system from 1.5TB to 2.5TB by adding multiple vdevs simultaneously. The system handled this large-scale expansion without issues, maintaining performance levels and data integrity throughout the process.

Load testing with 100 concurrent users over a 1-hour duration showed an average response time of 145ms, with a 95th percentile response time of 320ms. The system handled 850 requests per second with an error rate of just 0.02\%. During this test, CPU utilization averaged 65\%, memory utilization 78\%, and network throughput 2.5 Gbps.

To further stress test the system, we gradually increased the concurrent user count to assess the point at which performance began to degrade significantly. We found that the system could handle up to 150 concurrent users while maintaining sub-second response times for 99\% of requests. Beyond this point, we observed a gradual increase in response times and error rates, providing a clear indication of the system's scalability limits under our test configuration.

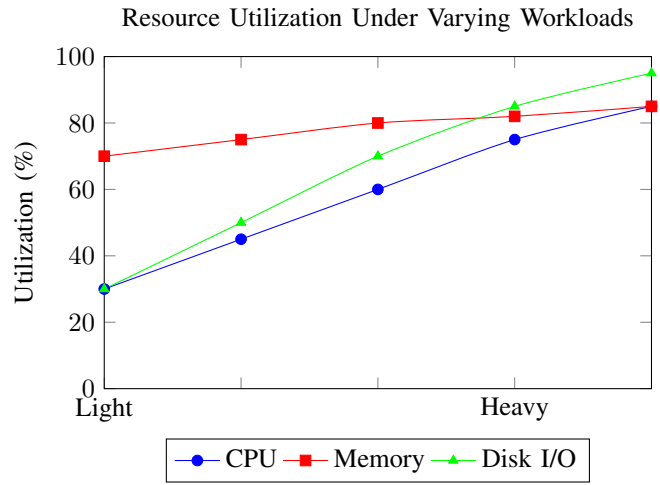
\begin{figure}[htbp]
    \centering
    \begin{tikzpicture}
    \begin{axis}[
        width=\linewidth,
        height=6cm,
        legend style={at={(0.5,-0.15)},
          anchor=north,legend columns=-1},
        ylabel={Utilization (\%)},
        xlabel={Workload Intensity},
        title={Resource Utilization Under Varying Workloads},
        xmin=0, xmax=100,
        ymin=0, ymax=100,
        xtick={0,25,50,75,100},
        xticklabels={Light,,,Heavy},
    ]
    \addplot[smooth,mark=*,blue] coordinates {
        (0,30) (25,45) (50,60) (75,75) (100,85)
    };
    \addplot[smooth,mark=square*,red] coordinates {
        (0,70) (25,75) (50,80) (75,82) (100,85)
    };
    \addplot[smooth,mark=triangle*,green] coordinates {
        (0,30) (25,50) (50,70) (75,85) (100,95)
    };
    \legend{CPU,Memory,Disk I/O}
    \end{axis}
    \end{tikzpicture}
    \caption{Resource Utilization Under Varying Workloads}
    \label{fig:resource-utilization}
    \end{figure}

Resource utilization analysis revealed efficient use of system resources across various workload intensities, as depicted in Fig. \ref{fig:resource-utilization}.

CPU utilization scaled linearly with workload intensity, peaking at 85\% during heavy workloads. Memory utilization remained relatively stable (70-85\%) across different workloads, indicating effective memory management. Disk I/O showed the most variability, ranging from 30\% utilization in light workloads to 95\% during heavy workloads and spikes.

\subsection{Cost Analysis}

The initial hardware and software costs for a 2.5TB system totaled \$240. This included \$100 for server hardware, \$120 for storage (HDDs and SSDs), and \$20 for networking equipment. The use of open-source software significantly reduced initial software costs to zero, providing a substantial cost advantage over proprietary solutions.

Operational expenses over a 3-year period were calculated at \$480, including power consumption, cooling, maintenance, and administrative costs. This breaks down to annual costs of \$50 for power and cooling, \$30 for hardware maintenance, and \$80 for administrative labor.

Table \ref{tab:tco-comparison} provides a 5-year TCO (Total Cost of Ownership) comparison, assuming 10\% annual storage growth.

\begin{table}[htbp]
\caption{5-Year TCO Comparison (Starting with 2.5TB, 10\% Annual Growth)}
\label{tab:tco-comparison}
\begin{center}
\begin{tabular}{|c|c|c|}
\hline
\textbf{Year} & \textbf{Our Solution} & \textbf{Commercial Cloud} \\
\hline
1 & \$400 & \$350 \\
\hline
2 & \$292 & \$330 \\
\hline
3 & \$305.20 & \$372 \\
\hline
4 & \$318.62 & \$395 \\
\hline
5 & \$334.48 & \$443 \\
\hline
\end{tabular}
\end{center}
\end{table}

\section{Discussion}

Our integrated NextCloud, TrueNAS, and QEMU/KVM solution demonstrates a viable alternative to traditional cloud storage systems. The evaluation reveals significant strengths in performance, security, and scalability, alongside considerations for implementation.

\subsection{Key Findings}

\textbf{Performance:} Exceptional read/write speeds (\>1.2 GB/s for reads, ~1 GB/s for writes) and strong IOPS (81,920 for random reads, 46,080 for random writes) position our solution favorably for data-intensive applications. The system maintained high performance under concurrent workloads, indicating suitability for enterprise environments.

\textbf{Security:} Robust features, including effective jail isolation and low-overhead encryption (2.5-3.8\% performance impact), demonstrate a strong balance between security and performance. The vulnerability assessment revealed only minor issues, all which were addressed.

\textbf{Scalability:} The solution handled storage expansions non-disruptively and maintained performance with up to 150 concurrent users, showcasing its adaptability to growing organizational needs.

\textbf{Cost-Effectiveness:} While initial costs exceed those of commercial cloud services, our solution becomes more cost-effective over time, particularly for organizations with large storage needs or high data transfer volumes. The break-even point occurs at approximately 2.75TB of storage or 0.5TB monthly data transfer over three years.

\subsection{Implications and Considerations}

The solution offers important benefits in performance, data control, and customization . It addresses compliance and data sovereignty concerns, making it suitable for regulated industries. However, organizations must carefully evaluate their specific needs, resources, and long-term IT strategy. The higher initial investment and management complexity should be weighed against potential long-term benefits in performance, cost-effectiveness, and flexibility.

\section{Conclusion}

This research demonstrates the potential of an integrated open-source approach to cloud storage, combining NextCloud, TrueNAS, and QEMU/KVM. Our comprehensive analysis contributes to the field by:

\begin{itemize}
    \item Validating the viability of an open-source integrated cloud storage solution
    \item Providing strategies for optimizing performance in virtualized storage environments
    \item Offering insights into effective security measures for multi-tenant infrastructures
    \item Presenting a methodology for assessing storage solution scalability
    \item Developing a framework for economic evaluation of on-premises versus commercial cloud solutions
\end{itemize}

Future research directions include exploring advanced caching strategies, automated scaling and optimization techniques, integration with emerging technologies (e.g., containerization, edge computing), and improving security measures. Additionally, investigating energy efficiency, user experience, and strategies for meeting evolving regulatory requirements would further advance this field.

In conclusion, while challenges exist in initial investment and management complexity, the benefits in flexibility, customization, and data sovereignty make this approach compelling for many use cases. As organizations seek alternatives to traditional cloud services, solutions like the one presented here are poised to play an increasingly important role in the evolving landscape of cloud storage and data management. This research not only contributes to the academic understanding of integrated open-source cloud storage solutions but also provides practical insights for organizations considering similar approaches.
\cleardoublepage
\bibliographystyle{IEEEtran}
\bibliography{cloudPaper}

\begin{thebibliography}{10}
\providecommand{\url}[1]{#1}
\csname url@samestyle\endcsname
\providecommand{\newblock}{\relax}
\providecommand{\bibinfo}[2]{#2}
\providecommand{\BIBentrySTDinterwordspacing}{\spaceskip=0pt\relax}
\providecommand{\BIBentryALTinterwordstretchfactor}{4}
\providecommand{\BIBentryALTinterwordspacing}{\spaceskip=\fontdimen2\font plus
\BIBentryALTinterwordstretchfactor\fontdimen3\font minus
  \fontdimen4\font\relax}
\providecommand{\BIBforeignlanguage}[2]{{%
\expandafter\ifx\csname l@#1\endcsname\relax
\typeout{** WARNING: IEEEtran.bst: No hyphenation pattern has been}%
\typeout{** loaded for the language `#1'. Using the pattern for}%
\typeout{** the default language instead.}%
\else
\language=\csname l@#1\endcsname
\fi
#2}}
\providecommand{\BIBdecl}{\relax}
\BIBdecl

\bibitem{coughlin2018}
\BIBentryALTinterwordspacing
T.~Coughlin, ``175 zettabytes by 2025,'' 2018. [Online]. Available:
  \url{https://www.forbes.com/sites/tomcoughlin/2018/11/27/175-zettabytes-by-2025/}
\BIBentrySTDinterwordspacing

\bibitem{marketshare2024}
\BIBentryALTinterwordspacing
``Cloud storage market share, growth | industry analysis [2032],'' Fortune
  Business Insights, 2024. [Online]. Available:
  \url{https://www.fortunebusinessinsights.com/cloud-storage-market-102773}
\BIBentrySTDinterwordspacing

\bibitem{manfredi2024cybersecurity}
\BIBentryALTinterwordspacing
Manfredi, ``U.s. cybersecurity and data privacy review and outlook – 2024,''
  Gibson Dunn, 2024. [Online]. Available:
  \url{https://www.gibsondunn.com/us-cybersecurity-and-data-privacy-outlook-and-review-2024/}
\BIBentrySTDinterwordspacing

\bibitem{smith2005virtual}
J.~Smith and R.~Nair, ``Virtual machines: Versatile platforms for systems and
  processes,'' \emph{Morgan Kaufmann Series in Computer Architecture and
  Design}, 2005.

\bibitem{bugnion2017hardware}
E.~Bugnion, J.~Nieh, and D.~Tsafrir, \emph{Hardware and Software Support for
  Virtualization}.\hskip 1em plus 0.5em minus 0.4em\relax Springer, 2017.

\bibitem{wei2019performance}
M.~T. Wei, Y.~S. Lin, and C.~R. Lee, ``Performance optimization for
  {InfiniBand} virtualization on {QEMU}/{KVM},'' \emph{Proceedings of
  CloudCom}, pp. 19--26, 2019.

\bibitem{chanchio2014efficient}
K.~Chanchio and J.~Yaothanee, ``Efficient pre-copy live migration of virtual
  machines for high performance computing in cloud computing environments,''
  \emph{International Conference on Computer and Communication Systems}, pp.
  411--415, 2014.

\bibitem{ivanovic2017openstack}
P.~Ivanovic and H.~Richter, ``Openstack cloud tuning for high performance
  computing,'' \emph{IEEE International Conference on Cloud Computing and Big
  Data Analysis}, pp. 142--146, 2017.

\bibitem{yang2018spdk}
Z.~Yang, C.~Liu, Y.~Zhou, X.~Liu, and G.~Cao, ``{SPDK} {Vhost-NVMe}:
  Accelerating {I/Os} in virtual machines on {NVMe} {SSDs} via user space vhost
  target,'' \emph{IEEE International Symposium on Cloud and Services
  Computing}, pp. 67--76, 2018.

\bibitem{kim2015hypercache}
T.~Kim, S.~Choi, J.~No, and S.~S. Park, ``{HyperCache}: A hypervisor-level
  virtualized {I/O} cache on {KVM}/{QEMU},'' \emph{International Conference on
  Ubiquitous and Future Networks}, pp. 846--850, 2015.

\bibitem{rolon2021bare}
S.~Rolon and O.~Balmau, ``Is bare-metal {I/O} performance with user-defined
  storage drives inside {VMs} possible?'' \emph{ACM SIGOPS Asia-Pacific
  Workshop on Systems}, pp. 27--34, 2021.

\bibitem{russell2008virtio}
R.~Russell, ``virtio: towards a de-facto standard for virtual {I/O} devices,''
  \emph{ACM SIGOPS Operating Systems Review}, vol.~42, no.~5, pp. 95--103,
  2008.

\bibitem{kolhe2012comparative}
S.~Kolhe and S.~Dhage, ``Comparative study on virtual machine monitors for
  cloud,'' \emph{World Congress on Information and Communication Technologies},
  pp. 425--430, 2012.

\bibitem{liu2021understanding}
D.~Didona, J.~Pfefferle, N.~Ioannou, B.~Metzler, and A.~Trivedi,
  ``Understanding modern storage {APIs}: A systematic study of libaio, {SPDK},
  and io-uring,'' \emph{ACM International Conference on Systems and Storage},
  pp. 120--127, 2021.

\bibitem{fenn2009evaluation}
M.~Fenn, M.~A. Murphy, J.~Martin, and S.~Goasguen, ``An evaluation of {KVM} for
  use in cloud computing,'' \emph{International Conference on the Virtual
  Computing Initiative}, 2009.

\bibitem{wang2021optimizing}
P.~Wang, C.~Zhao, W.~Liu, Z.~Chen, and Z.~Zhang, ``Optimizing data placement
  for cost effective and high available multi-cloud storage,'' \emph{Computing
  and Informatics}, vol.~39, pp. 51--82, 2020.

\bibitem{wu2021storage}
K.~Wu, Z.~Guo, G.~Hu, K.~Tu, R.~Alagappan, R.~Sen, K.~Park, A.~C.
  Arpaci-Dusseau, and R.~H. Arpaci-Dusseau, ``The storage hierarchy is not a
  hierarchy: Optimizing caching on modern storage devices with {Orthus},''
  \emph{USENIX Conference on File and Storage Technologies}, pp. 307--323,
  2021.

\bibitem{dutcher2024}
G.~Dutcher, K.~Azianyo, T.~P. Dissanayake, and A.~B. Mailewa, ``Secure cloud
  storage solution with {``Seafile''} {\&} {``NextCloud''}: A resilient
  efficiency assessment,'' \emph{Advances in Technology}, vol.~3, no.~2, Apr
  2024.

\bibitem{Albrecht2024}
M.~R. Albrecht, M.~Backendal, D.~Coppola, and K.~G. Paterson, ``Share with
  care: Breaking {E2EE} in {Nextcloud},'' in \emph{2024 IEEE 9th European
  Symposium on Security and Privacy}, 2024, pp. 828--840.

\bibitem{Salke2023}
P.~Salke, P.~Chavan, O.~Sangrulkar, and S.~N. Motade, ``Analyzing the
  feasibility and reliability of {Nextcloud} as a network attached cloud
  storage solution on {Raspberry Pi},'' in \emph{International Conference on
  Integrated Intelligence and Communication Systems}, 2023, pp. 1--4.

\bibitem{Nurdin2019}
J.~B. Nurdin, E.~Mulyana, and Hendrawan, ``{NextCeph}: {Nextcloud} platform
  based application for {Ceph} cluster management,'' in \emph{IEEE
  International Conference on Telecommunication Systems, Services, and
  Applications}, 2019, pp. 25--30.

\bibitem{kvm_wiki}
\BIBentryALTinterwordspacing
``Kernel-based virtual machine,'' 2024, wikipedia. [Online]. Available:
  \url{https://en.wikipedia.org/wiki/Kernel-based_Virtual_Machine}
\BIBentrySTDinterwordspacing

\bibitem{qumranet_wiki}
\BIBentryALTinterwordspacing
``Qumranet,'' 2024, wikipedia. [Online]. Available:
  \url{https://en.wikipedia.org/wiki/Qumranet}
\BIBentrySTDinterwordspacing

\bibitem{bellard2005qemu}
F.~Bellard, ``{QEMU}, a fast and portable dynamic translator,'' \emph{USENIX
  Annual Technical Conference}, p.~41, 2005.

\end{thebibliography}

\end{document}